\documentclass[aps, prl,longbibliography, 
amsmath,amssymb,
 reprint,%
]{revtex4-2}

\usepackage{graphicx}
\usepackage{dcolumn}
\usepackage{dsfont}
\usepackage{bm}

\usepackage[T1]{fontenc}
\usepackage{etoolbox}
\usepackage{xspace}
\usepackage{array}
\usepackage{siunitx}
\usepackage{booktabs}
\usepackage{bigstrut}
\usepackage{stmaryrd}

\usepackage{longtable} 
\usepackage{tabularx} 

\newcolumntype{K}[1]{>{\centering\let\newline\\\arraybackslash\hspace{0pt}}m{#1}}
\usepackage{makecell}
\usepackage{float}

\usepackage[hidelinks]{hyperref}
\hypersetup{
  colorlinks   = true, 
  urlcolor     = black, 
  linkcolor    = black, 
  citecolor   = black 
}
\usepackage{soul}

\makeatletter
\def\@email#1#2{%
 \endgroup
 \patchcmd{\titleblock@produce}
  {\frontmatter@RRAPformat}
  {\frontmatter@RRAPformat{\produce@RRAP{*#1\href{mailto:#2}{#2}}}\frontmatter@RRAPformat}
  {}{}
}%

\usepackage[mathscr]{euscript}
\usepackage{color}
\usepackage{xcolor}
\usepackage{amsmath}
\usepackage[bb=boondox]{mathalfa}
\usepackage{mathrsfs}
\usepackage{multirow}
\usepackage{braket}
\usepackage{mathtools}

\newcommand{\fumarateC}{[1--\textsuperscript{13}C]-fumarate\,}

\newcommand{\icfo}{ICFO—Institut de Ciències Fotòniques, The Barcelona Institute of Science and Technology, Castelldefels
(Barcelona) 08860, Spain.}
\newcommand{\ibec}{IBEC -- Institute for Bioengineering of Catalonia, 08028 Barcelona, Spain.}
\newcommand{\fzj}{Institute of Biological Information Processing, Structural Biochemistry (IBI-7), Forschungszentrum J\"ulich, 52428 J\"ulich, Germany.}
\newcommand{\rwth}{Institute of Technical and Macromolecular Chemistry, RWTH Aachen University,  52074 Aachen, Germany.}
\newcommand{\dwi}{DWI -- Leibniz-Institute for Interactive Materials, 52074 Aachen, Germany.}
\newcommand{\iut}{Department of Electrical and Computer Engineering, Isfahan University of Technology, 84153-83111 Isfahan, Iran.}
\newcommand{\icrea}{ICREA -- Instituci\'{o} Catalana de Recerca i Estudis Avan\c{c}ats, 08010 Barcelona, Spain.}

\makeatother
\begin{document}

\preprint{APS/123-QED}

\title{Dynamics of hyperpolarized nuclear spins 
at Zeeman-insensitive points}

\author{James Eills\textsuperscript{1,2}}
\author{Anushka~Singh\textsuperscript{2,3,4}}
\author{Amir Mahyar Teimoori\textsuperscript{5,6}}
\author{Irene Marco-Rius\textsuperscript{1}}
\author{Morgan W. Mitchell\textsuperscript{5,7}}
\author{Michael C. D. Tayler*\textsuperscript{5}}
\email{michael.tayler@icfo.eu}

\affiliation{\textsuperscript{1} \ibec}
\affiliation{\textsuperscript{2} \fzj}
\affiliation{\textsuperscript{3} \rwth}
\affiliation{\textsuperscript{4} \dwi}
\affiliation{\textsuperscript{5} \icfo}
\affiliation{\textsuperscript{6} \iut}
\affiliation{\textsuperscript{7} \icrea}

\date{\today}

\begin{abstract}
Avoided crossing (LAC) transitions are central to spin-based precision measurements since they provide strong observability and field-perturbation insensitivity, but while observed commonly in isolated-atom or engineered solid-state platforms, such transitions are unexploited in molecules in liquids.
Here we report a solution-state, nuclear-spin LAC transition in hyperpolarized \fumarateC. 
The singlet--triplet LAC, occurring near $\SI{400}{\nano\tesla}$, produces pronounced magnetization oscillations at approximately $\SI{2}{Hz}$ with a coherence time of $\SI{25}{s}$, three times longer than the longest single-spin $T_2^*$. 
In the regime of high spin concentration and high polarization the sample's internal dipolar field couples measurably to the transition, providing a direct probe of nonlinear spin dynamics. 
These results establish nuclear-spin LACs as a platform for quantum sensing in liquids.
\end{abstract}
                              
\maketitle

\noindent Nuclear spin hyperpolarization\cite{eills2023spin}  unites magnetization far beyond thermal equilibrium with the intrinsic advantages that make nuclear spins powerful sensors: long lifetimes, tunable coupling topologies and robust control.  
While many applications including NMR and MRI have advanced significantly due to these combined benefits, others remain comparatively underdeveloped, including searches for exotic spin couplings \cite{Chu2020,Arvanitaki2014} (e.g., axion-like pseudo-fields), and the use of long-range dipolar fields to access nonlinear magnetization dynamics \cite{Warren1993,Ledbetter2002}. 

The key requirement is a measurement setting that separates the exotic dynamics from the noisy magnetic-field environment, Hyperpolarized atoms, \textit{e.g.}, \textsuperscript{3}He \cite{Baudin2007} or \textsuperscript{129}Xe \cite{Limes2018}, are the current leading platform due to their long coherence times across condensed phases.  Hyperpolarized molecules, in comparison, lie far behind despite multiple known approaches to mitigate field noise, such as 
uniform, fluctuation-free magnets \cite{bax_enhanced_1980,heil_spin_2013,lacey_high-resolution_1999}, 
dynamical decoupling \cite{Sahin2022}, 
comagnetometry \cite{Ledbetter2012PRLcomagnetometer,Wu2018comagnetometer,Garcon2019} and 
indirect zero-quantum (ZQ) spectroscopy  \cite{aue_twodimensional_1976,munowitz_multiple-quantum_1986,vathyam_homogeneous_1996,cai_high-resolution_2009,pelupessy_high-resolution_2009,terenzi_enabling_2019}.  
These methods, moreover, do not always match with single-shot hyperpolarization methods and real-time coherent control.

Here we report a different approach: detection at a level anticrossing (LAC), where observability and field insensitivity coexist.
The LAC provides a discrete operating field where an observable transition has a leading order frequency vs.\ field derivative of zero \cite{eck_level_1967}; small changes in field are consequently negligible.  
The convenience of such operating points in precision sensing is exemplified across many diverse electro-optical platforms, such as microwave atomic clocks, rare-earth quantum memories \cite{wang2025nuclear} and atom-vacancy magnetometers \cite{broadway2016anticrossing}, as well as superconducting qubit circuits \cite{manucharyan2009fluxonium,bao2022fluxonium}.

\begin{figure}
    \centering
    \includegraphics[width=\columnwidth]{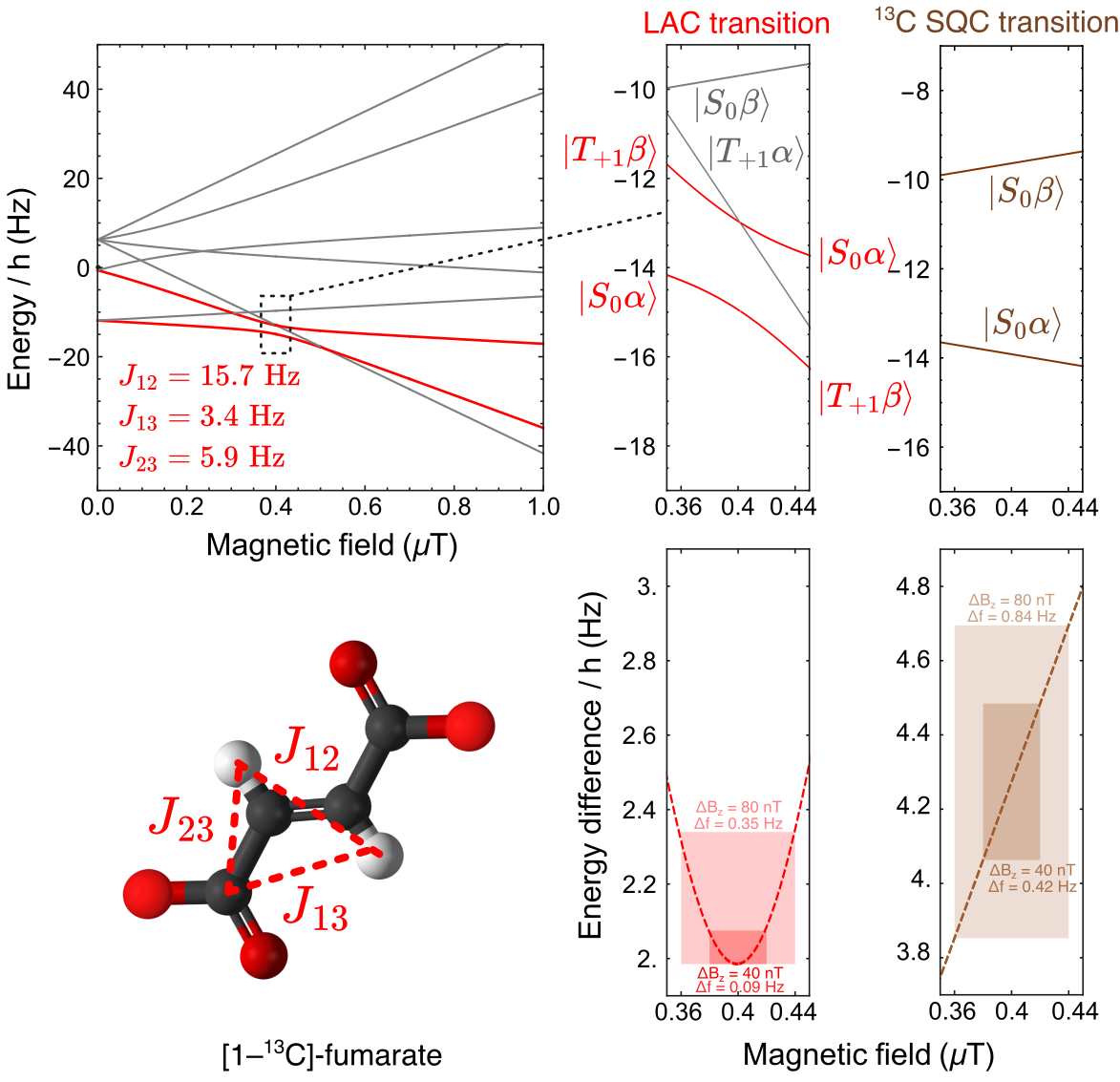}
    \caption{Comparison of magnetic field-to-frequency gradient between the observable LAC transition in \fumarateC, and a \textsuperscript{13}C single-quantum (SQC) transition in a hypothetical non-coupled molecule, where $J_{13} = J_{23} = 0$ Hz.}
    \label{fig:fig1}
\end{figure}

To demonstrate the utility of Zeeman-insensitive points 
to probe hyperpolarization dynamics, we use \fumarateC, a three-spin molecule in which two \textsuperscript{1}H and one \textsuperscript{13}C nuclei are scalar coupled and exhibit a LAC below Earth's field, near \SI{400}{\nano\tesla} (\autoref{fig:fig1}). 
We obtain orders-of-magnitude signal enhancement by hyperpolarizing \fumarateC prior to detection.
This anticrossing is already central to parahydrogen-induced hyperpolarization (PHIP) of fumarate, where adiabatic passage through the LAC can be used to convert parahydrogen singlet order into net nuclear magnetization \cite{johannesson2004transfer}. 
The same mechanism enables highly polarized fumarate for metabolite imaging in vitro \cite{eills_real-time_2019} and in vivo \cite{gallagher_production_2009,gierse_parahydrogen-polarized_2023}; thus, the LAC already sits in an application-relevant regime.

\textit{Background-field insensitivity --- } The LAC in \fumarateC is understood from the Hamiltonian of its scalar-coupled nuclei.
Although the full Hilbert space is $2^3=8$ dimensional, conservation of total projection angular momentum ($m_I$) partitions it into four independent subspaces: dimension 1, 3, 3 and 1 for $m_I=-3/2,-1/2,+1/2$ and $+3/2$, respectively. 
The relevant LAC occurs in one of the $m_I=\pm1/2$ manifolds; a predominantly \textsuperscript{1}H singlet state, e.g.,  in $m_I=+1/2$, 
\begin{subequations}
\begin{equation}
\ket{\phi_a} =  \ket{S_0\alpha} \equiv (\ket{\alpha\beta\alpha}-\ket{\beta\alpha\alpha})\sqrt{2} 
\label{eq:defnS0a}
\end{equation}
(notation: $\ket{\alpha\beta\alpha}\equiv \ket{\alpha_1} \otimes \ket{\beta_2} \otimes \ket{\alpha_3}$, for indices $1,2=$\textsuperscript{1}H, $3=$\textsuperscript{13}C) 
is brought into resonance with a triplet-like state
\begin{equation}
\ket{\phi_b} = \ket{T_{+1}\beta} \equiv \ket{\alpha\alpha\beta},\label{eq:defnT1b}
\end{equation}
while both are only weakly coupled to the third state
\begin{equation}
\ket{\phi_c} = \ket{T_0\alpha} \equiv (\ket{\alpha\beta\alpha}+\ket{\beta\alpha\alpha})\sqrt{2}\,.
\label{eq:defnT0a}
\end{equation}
\end{subequations} 
As Appendix A shows, the LAC occurs at a field 
\begin{equation}
    B_{\rm LAC} \approx  {\rm{sign}}(m_I) \frac{\pi(4J_{12}-J_{13} - J_{23})}{2(\gamma_1-\gamma_3)}
    \,. \label{eq:bprimedef}
\end{equation}
which from the scalar couplings $J_{12}=$ \SI{15.70}{Hz}, $J_{13}=$ \SI{5.94}{Hz} and $J_{23}=$ \SI{3.40}{Hz} for \fumarateC, is expected at $B_{\rm LAC} \approx +$\SI{400}{nT} \cite{rodin2021constant,eills_live_2024}.  The LAC of the $m_I=-1/2$ manifold occurs at negative field.
Appendix A also shows that the leading Hamiltonian at $B_{\rm LAC}$ is proportional to the unity operator of the $m_I$ subspace; thus, the LAC is first-order insensitive to magnetic field changes.  The energy splitting between $\ket{\phi_a}$ and $\ket{\phi_b}$ is set predominantly by the difference in the \textsuperscript{1}H--\textsuperscript{13}C couplings,
\begin{equation}
 \omega_{\rm LAC} = \pi\sqrt{2}\,|J_{13}-J_{23}| \,,\label{eq:wlaczeroorder}
\end{equation}
or $|J_{13}-J_{23}|\sqrt{2} \equiv J_{\Delta}/\sqrt{2}$ in Hz.
A second-order correction to include the non-secular coupling to $\ket{\phi_c}$ gives $\omega_{\rm LAC} = \pi\sqrt{2}(1 - x + x^2)|J_{13}-J_{23}|$, where $x = (J_{13}+J_{23})/ (4J_{12})$.

Expanding about $B_{\rm LAC}$, with $\delta B = B_z - B_{\rm LAC}$, gives the quadratic dependence 
\begin{eqnarray}
    \omega_{\rm LAC}(\delta B) 
    &=& 
    \omega_{\rm LAC}(0) + \frac{(\gamma_{\rm H} - \gamma_{\rm C})^2 \delta B^2}{2\pi\sqrt{2}\, J_\Delta} + O(\delta B^4)\,,\label{eq:wlac2ndorder}
\end{eqnarray}
and the curvature determines the usable field-insensitive window.  Comparing the local slope of ${\rm d}\omega_{\rm LAC}/{\rm d}B_z$ to the Zeeman slope of the slowest observable single-quantum (SQ) transition,  
$|{\rm d}\omega_{\rm SQ}/{\rm d}B_z| = \gamma_{\rm C}$, we see that the LAC is less sensitive to inhomogeneous broadening over a field interval of order $(J_{13} - J_{23})/\gamma_{\rm H}$.  This range is approximately \SI{350}{nT} to \SI{450}{nT} as illustrated in \autoref{fig:fig1}.

In contrast, effective or pseudo fields\cite{ledbetter2012liquid} that couple to nuclear spins independently of $\gamma_{\rm{H}}$ and $\gamma_{\rm{C}}$ are not canceled by the LAC condition, and  generate a first-order frequency shift. 
This provides the sensitivity to exotic couplings, including the axion-like fields targeted in CASPEr-ZULF NMR experiments \cite{Garcon2019}.

\begin{figure*}
    \centering
    \includegraphics[width=0.8\textwidth]{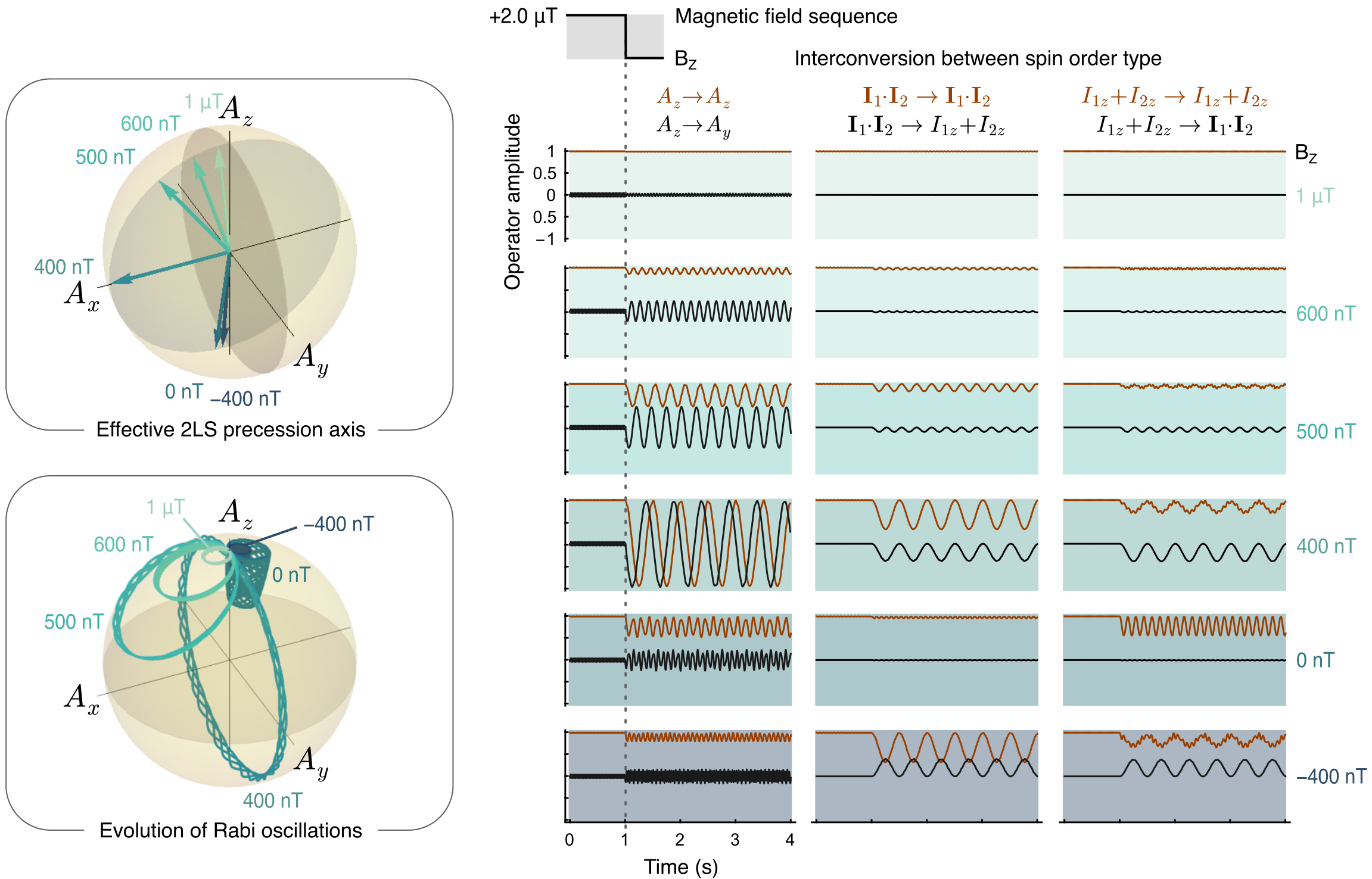}
    \caption{Magnetization evolution near $B_{\rm LAC}$: (\textit{Left}) Vector representations of the $\{\ket{\phi_a},\ket{\phi_b}\}$ subspace in \fumarateC, showing the time-evolution of $\rho^{\rm(ab)}(0)=A_z$ reflects precession about the effective axis (\autoref{eq:3DsubspaceLvN}+\autoref{eq:wlacvectormodel}).
    (\textit{Right}):
    Simulated spin trajectories when switching from an initial field $B_z=$ \SI{2}{\micro\tesla} to a lower field at \SI{1}{s}, also for $\rho^{\rm(ab)}(0)=A_z$.  
    $I_{1z}+I_{2z}$ is proportional to the \textsuperscript{1}H $z$-axis magnetization, and $\mathbf{I}_1\cdot\mathbf{I}_2$ is the traceless \textsuperscript{1}H spin-singlet order.  Conversion is most efficient at $B_z=\pm B_{\rm LAC}$, reducing as field offset increases. Oscillations at $B_z=\pm B_{\rm LAC}$ are equal in amplitude but opposite in sign, due to the same initial population of the $m=\pm1/2$ manifolds.  Switching to zero field yields no net magnetization.}
    \label{fig:fig2}
\end{figure*}

\textit{Magnetization dynamics --- } The LAC transition is directly observable because population motion between $\ket{\phi_a}$ and $\ket{\phi_b}$ changes the net magnetic moment of the molecule along the laboratory $z$ axis.  
As Appendix B shows, the dynamics can be described by a three-dimensional Bloch vector 
$\mathbf{\rho}^{\rm(ab)}(t)$
with components along effective operators $A_x, A_y, A_z$. Operator $A_z$ describes population imbalance between $\ket{\phi_a}$ and $\ket{\phi_b}$, while operators $A_x$ and $A_y$ are corresponding coherences. 
The projected dynamics are a Bloch-like precession
\begin{equation}
\bm{\rho}^{\rm(ab)}(t) \approx R{(
 \bm{\omega}_{\rm LAC}\, t)}\bm{\rho}^{\rm(ab)}(0)\,,  \label{eq:3DsubspaceLvN}  
\end{equation}
about the effective angular velocity vector
\begin{eqnarray}
 \bm{\omega}_{\rm LAC} 
 &=& (
 \pi\sqrt{2}J_\Delta,\,0,\,(\gamma_{\rm H}-\gamma_{\rm C})\delta B)\,, \label{eq:wlacvectormodel}
\end{eqnarray}
and this motion is illustrated in \autoref{fig:fig2}.  At the LAC, $\bm{\rho}^{\rm(ab)}$ precesses about the $A_x$ axis, while away from the LAC the precession axis tilts towards $\rm sign(\delta B) A_z$.

Also explained in Appendix B, a change in polar angle of the Bloch vector generates a change in magnetization of the molecule.  The resulting moment, $M_z$, is along the laboratory $z$ axis.  

We now explicitly relate this to fumarate and analogous AA'X molecules produced in hyperpolarized form by PHIP reactions.  The ideal conversion is to a pure singlet-polarized spin system, corresponding here to a density operator $\rho(0) = \ket{S_0\alpha}\bra{S_0\alpha} + \ket{S_0\beta}\bra{S_0\beta} \equiv  1/4 -\mathbf{I}_1\cdot\mathbf{I}_2$, or equivalently $\mathbf{\rho}^{\rm(ab)}(0) = (0,0,0.5)$ in both the $m=+ 1/2$ and $m=- 1/2$ subspaces.
Sudden evolution at $\delta B=0$ then equates to precession in the $yz$ plane, and the $z$-axis magnetization follows
\begin{eqnarray}
    \braket{M_z}(t) &\propto& \frac{\gamma_{\rm H} - \gamma_{\rm C}}{4} \Big[\cos(\theta(t)) -1\Big]\,,\label{eq:MzvsTime}
\end{eqnarray}
where $\theta(t) = \omega_{\rm LAC} t$ is the time-dependent polar angle.  As $\braket{M_x} = \braket{M_y} = 0$, the total magnetization oscillates along $z$ from zero to a maximum proportional to $(\gamma_{\rm H} - \gamma_{\rm C})$, (see \autoref{fig:fig2}, center+right), and the heteronuclear nature of the system is key to observability. 
For nonzero $\delta B$, evolution is constrained by $\theta(t)\leq \theta_{\rm max}$ where
\begin{eqnarray}
    |\theta_{\rm max}| &\leq& 
    2 \arctan\!\left|\frac{\delta B(\gamma_{\rm H}-\gamma_{\rm C})}{\pi\sqrt{2}J_\Delta}\right|\,.
    \label{eq:17}
\end{eqnarray}

\begin{figure*}
    \centering
    \includegraphics[width=0.7\textwidth]{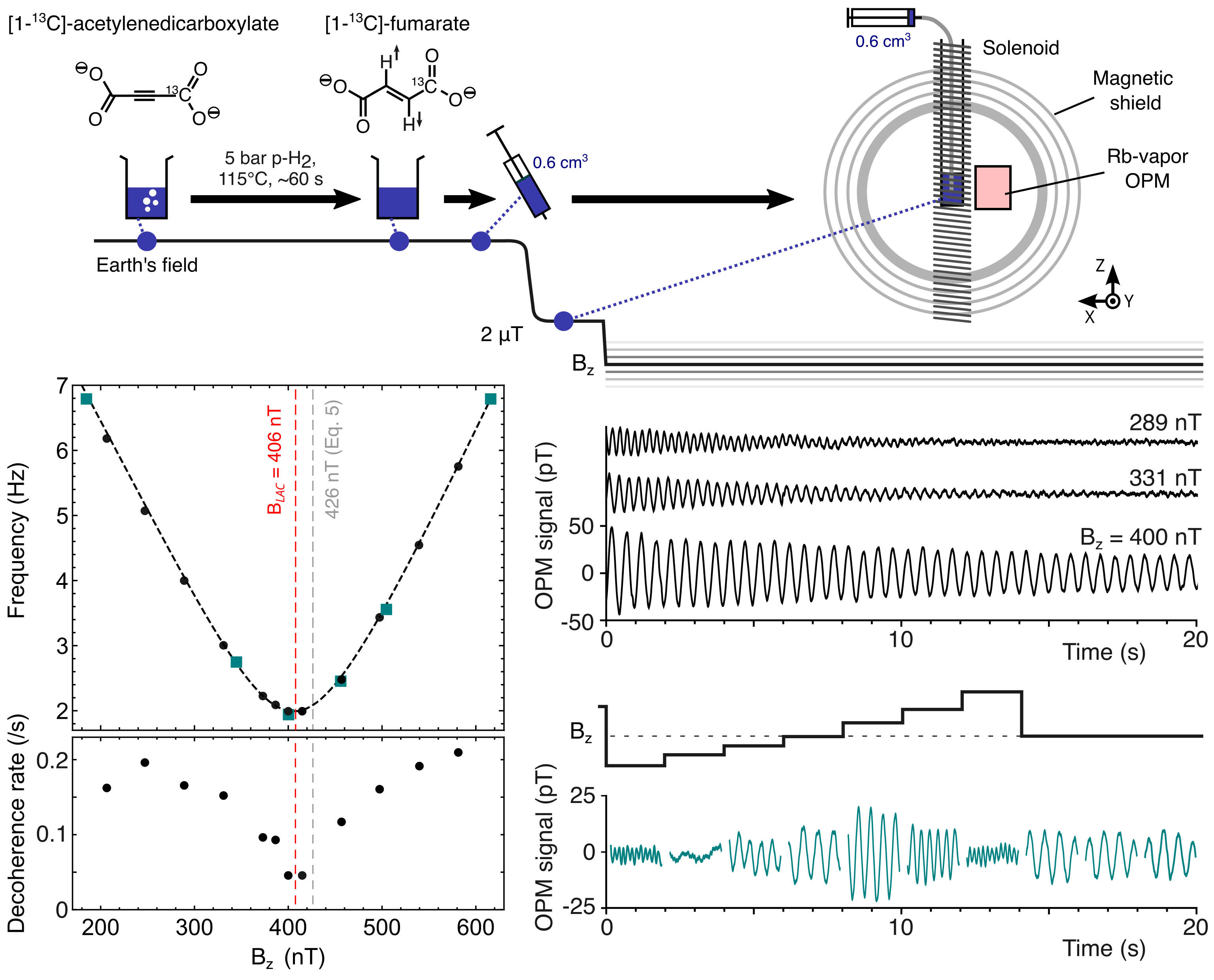}
    \caption{Solution-state LAC coherences in hyperpolarized \fumarateC observed via magnetometry. 
    \textit{(Top)}: Experimental scheme, showing \fumarateC initialized in a near-pure \textsuperscript{1}H-singlet state by PHIP, then syringe transferred to a low-field setup for field-switching and measurement of the nuclear magnetization using an OPM.
    (\textit{Bottom}): Experimental magnetization curves, where colors indicate: (\textit{Black}) multiple samples, each with a single switch from \SI{2}{\micro\tesla} to a different $B_z$; (\textit{Teal}) a single sample with switching every \SI{2}{s} to multiple different $B_z$ fields.
   Plots to the left show the fitted frequency and decoherence rates \textit{vs.}\ $B_z$. On the frequency plot, the black dashed curve is the calculated eigenvalue difference between the evolution states of the Hamiltonian in \autoref{eq:HBz}, with best-fit coupling parameters $J_{12} = \SI{15.92}{\hertz}$, $J_{\Sigma} = \SI{9.34}{\hertz}$ and $J_{\Delta} = \SI{2.514}{\hertz}$. Dashed red and gray lines indicate the values of $B_{\rm LAC}$ from exact diagonalization of \autoref{eq:HBz}, and from \autoref{eq:bprimedef}, respectively.}
    \label{fig:fig3}
\end{figure*}

\begin{figure*}
\includegraphics[width=\textwidth]{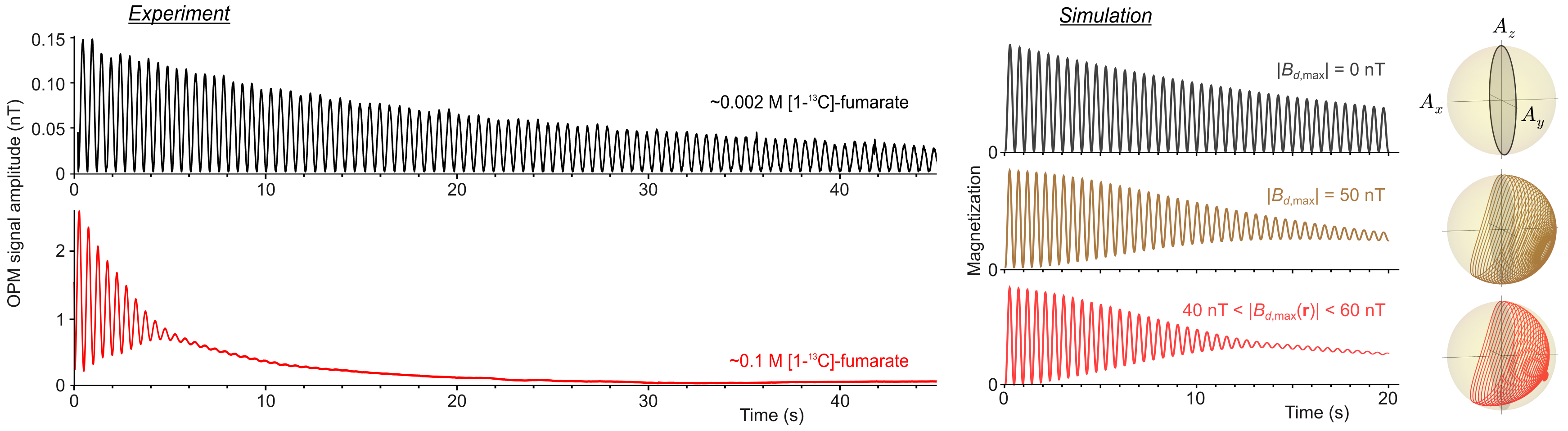}
    \caption{Dynamics in the high-magnetization regime.
    (\textit{Left}) Experimental magnetization trajectories of the LAC coherences for two samples with the same fumarate concentration, but different \textsuperscript{13}C isotope fraction. The vertical axis corresponds to magnetic field measured at the OPM, which is around 50 times smaller than the sample's internal dipolar field $B_d$.
    (\textit{Right}) A model representation of the high magnetization regime through simulated trajectories of the $\mathbf{\rho}^{(ab)}$ Bloch vector and magnetization projection $\braket{M_z(t)}\propto (1-\braket{A_z(t)}) /2$.  Each row shows evolution from the same initial condition, $\rho^{(ab)}(0) = A_z$, $\omega_{\rm LAC}/(2\pi) = \SI{2}{Hz}$, for different fields: 
    (\textit{top}) a constant on-resonance field, $\delta B = 0$;
    (\textit{center}) a time-varying pseudo-field due to the macroscopic magnetization $\delta B(t) = B_{d,\rm max} \times \braket{M_z(t)}$, with uniform $B_{d,\rm max} = \SI{50}{nT}$;
    (\textit{bottom}) an average of trajectories with time-dependent $\delta B$, across a uniform distribution of $B_{d,\rm max}$ between $\SI{40}{nT}$ and $\SI{60}{nT}$. }
\label{fig:fig4}
\end{figure*}

\textit{Experimental observation ---} \autoref{fig:fig3} shows the magnetization oscillations of fumarate at natural \textsuperscript{13}C isotopic abundance (2.2\% \textsuperscript{13}C) after hyperpolarization to a state of near-pure \textsuperscript{1}H spin-singlet order.  In the scheme shown, a precursor chemical acetylene dicarboxylate is combined with para-H\textsubscript{2} for $\sim$60 s in the ambient laboratory field to generate hyperpolarized fumarate at 60\,mM in D\textsubscript{2}O.
After cooling and depressurization, the solution is transferred to a magnetically shielded solenoid with field $B_z = \SI{2}{\micro\tesla}$. In this location the longitudinal magnetization is monitored by an alkali-vapor magnetometer \cite{eills_live_2024,mouloudakis2023real}.  
Overall this procedure is uncomplicated and inexpensive. 
The magnetometer is sensitive to sub-pT changes in the magnetic moment of the sample and produces no measurable effect on the nuclear spins being studied.  
This offers further simplicity compared to high-field NMR where inductive detection leads to complex dynamics via coil back-action or radiation damping \cite{krishnan2013radiation}.  


Sub-ms switching of $B_z$ to near $B_{\rm LAC}$ produces signals like those shown by the black curves in the center column of \autoref{fig:fig2}.  Excellent model-vs.-experiment agreement is given by performing a series of polarization and detection runs switching to different fields -- see \autoref{fig:fig3} for evolution fields of 289, 331 and 400\,nT -- and observing that the global fitted-frequency vs.\ $B_z$ curve is parabolic near the minimum and turning linear at offsets beyond $\pm$\SI{100}{nT}.
Fitting the eigenvalues of $H$ to the data points gives the dashed curve and refined values of the molecular $J$-couplings, $J_{12} = \SI{15.92}{\hertz}$ and $J_{\Sigma} = \SI{9.34}{\hertz}$, (which are mostly pinned down by $B_{\rm LAC}$), and $J_{\Delta} = \SI{2.514}{\hertz}$ (mostly pinned down by $\omega_{\rm LAC}$).  
The fitted minimum is $B_z=$ \SI{406(2)}{nT} and $\omega_{\rm LAC}/(2\pi) =$ \SI{1.995(5)}{Hz}. 
These values of $B_{\rm LAC}$ and $\omega_{\rm LAC}$ are around \SI{20}{nT} and \SI{0.22}{Hz} lower, respectively, than the predictions of \autoref{eq:bprimedef} and \autoref{eq:wlaczeroorder} using the same $J$-coupling values. 
The discrepancy arises due to the non-secular Hamiltonian terms that couple $\{\ket{\phi_a},\ket{\phi_b}\}$ with $\ket{\phi_c}$, but are accounted for by including the $J_\Sigma/(4J_{12})$ correction term.

A variant of this experiment was also performed using a single hyperpolarized sample, where $B_z$ was switched to a different value every \SI{2}{s} to trace out the frequency curve in a one-shot experiment (\autoref{fig:fig3}, lower right).  This method also gives excellent estimation of $B_{\rm LAC}$, despite the short measurement time at each field point.  

\textit{Coherence decay ---}
Within $\delta B =\pm  \SI{150}{nT}$ of $B_{\rm LAC}$ the observed coherence decay rate varies by close to a factor of 3.  The minimum at around $\delta B=0$ is, of course, the defining feature of the LAC point where a monoexponential decay time constant of \SI{25(1)}{s} is reached.  
This is substantially longer than the $T_2^*$ of any individual spin in the molecule under similar conditions, including the \textsuperscript{13}C SQC relaxation time; a pulse-acquire experiment at $B_z = \SI{2}{\micro\tesla}$ for the same solution gives $T_2^*(\textsuperscript{13}\rm C) = \SI{8.5(5)}{s}$.

Around 150-\SI{200}{nT} away from $B_{\rm LAC}$ the fitted decay rate was nearly constant, and this is consistent with inhomogeneous broadening as the dominant dephasing mechanism.  Away from the LAC the transition frequency should vary linearly with $B_z$, so a fixed field distribution ($\Delta B$) should produce a fixed spread of frequency ($\Delta f$) across the sample ($|\Delta B / \Delta f| = |2\gamma_{\rm H} - \gamma_{\rm C}| / (2\pi) $). \autoref{fig:fig1} supports the interpretation by showing that the spectrum is indeed linear in these ranges.

\textit{Nonlinearity at high magnetization ---}
Using a precursor chemical with 99\% $[^{13}\mathrm{C}_1]$ enrichment, we obtain PHIP-polarized \fumarateC \,in roughly 45 times higher concentration.  Here, a repeat of the sudden-switch experiment yields magnetization oscillation along $z$ at the same frequency but a significantly faster decay, within $\sim\SI{5}{s}$, with a distinctly non-exponential profile (see \autoref{fig:fig4}).
After the oscillating signal subsides a non-oscillating component persists, which relaxes exponentially with a decay time constant of 7 to \SI{8}{s}.

This faster decay of the oscillatory signal for the enriched sample cannot be attributed increased homogeneous relaxation because the total concentration of fumarate remains constant.  More likely, it is due to inhomogeneity generated by the liquid itself, since the PHIP-polarized fumarate now produces a substantial dipolar magnetization, $\mathbf{M}(\mathbf r)$ \cite{levitt1996demagnetization}
The mean self-generated dipolar field $\mathbf{B}_{\rm d}(\mathbf r) \propto \mathbf{M}(\mathbf r)$ is measured to be around $|B_{\mathrm d}|\sim\SI{50}{nT}$ from the \textsuperscript{1}H Larmor frequency shift of the sample in its own field \cite{eills_live_2024}.

This shift adds a substantial nonlinearity by making $\delta B$ time-dependent.  Given \autoref{eq:wlacvectormodel},  the effective field lies parallel to $A_x$ for a non-magnetized sample at the LAC point but around \SI{40}{\degree} out of plane at point of highest magnetization.  
\autoref{fig:fig4} simulates this effect, which is essentially the nuclear spin analog of a Josephson oscillation[], or self-locking between the magnetization and mean dipolar field axes.
Inhomogeneity across the sample volume, which is in practice certain, lowers the oscillation time.

Such effects and possible mitigation strategies, e.g., dipolar field decoupling, may also be investigated in other molecules.  Maleic acid dimethyl ester is one candidate \cite{buljubasich2012level}, which can be hyperpolarized via PHIP at room-temperature and because of its higher solubility produce demagnetizing fields at least an order of magnitude higher than achieved here in fumarate \cite{dagys2024robust}.
Comparable internal magnetization fields and nonlinear dynamics have previously been studied only under extreme conditions, such as in liquid \textsuperscript{3}He, \textsuperscript{129}Xe in ultrastrong magnets. 

In summary, we demonstrate a directly observable zero first-order Zeeman-insensitive transition in a liquid-state environment, without strong magnetic fields or low temperatures, which is readily hyperpolarized and sensitive to nonlinear spin dynamics.
To our best knowledge this is the first studied for nuclear spins in a molecule.
Similar $B_z$-inhomogeneity protected transitions should exist in many other molecules, since a magnetically active LAC should exist in any heteronuclear-coupled spin system. Hyperpolarization-relevant systems should include AA'XX' coupled molecules, such as when the \textsuperscript{1}H\textsubscript{2} pair of parahydrogen combines with \textsuperscript{13}C\textsubscript{2}- or \textsuperscript{15}N\textsubscript{2} molecules in either hydrogenative \cite{ivanov2014role} or exchange-mediated solution-state PHIP \cite{theis2015microtesla}.

\section{End Matter}
\textit{Appendix A. Hamiltonian near the anticrossing.---} 
The behavior of \fumarateC  near the low-field LAC follows from the spin Hamiltonian of the molecule in solution,
\begin{equation}
H=\sum_{j=1}^{3}\!\left[-\gamma_j B_z I_{jz}
+2\pi\!\!\sum_{k<j}J_{jk}\,\mathbf{I}_j\!\cdot\!\mathbf{I}_k\right],   \label{eq:Htotal}
\end{equation}
where $j=1,2$ label the protons and $j=3$ labels the \textsuperscript{13}C nucleus, $J_{jk}$ are scalar couplings in Hz, and $H$ is written in angular frequency units.  The applied field is $\textbf{B}=(0,0,B_z)$.

In the basis $\{\ket{\phi_a},\ket{\phi_b},\ket{\phi_c}\}$, defined by \autoref{eq:defnS0a}--\autoref{eq:defnT0a}
the matrix representation of $H$ shows states are connected by the respective sum and difference of the heteronuclear pair couplings. Defining $J_\Sigma = J_{13}+J_{23}$ and $J_\Delta = J_{13}-J_{23}$, we obtain
\begin{eqnarray}
H_{B_z} &\sim&
\begin{bmatrix}
\begin{array}{c|c|c}
 - 3J_{12} & -J_\Delta \sqrt{2} & J_\Delta \\ \hline
 -J_\Delta \sqrt{2} & J_{12} - J_\Sigma & J_\Sigma \sqrt{2} \\\hline
J_\Delta & J_\Sigma \sqrt{2} & J_{12}
\end{array}
\end{bmatrix}
\Bigr(\frac{\pi}{2}\Bigl) \nonumber \\&& -
\begin{bmatrix}
\begin{array}{c|c|c}
\gamma_3 & 0 & 0 \\\hline
0 & \gamma_1 + \gamma_2 -\gamma_3 & 0 \\ \hline
0 & 0 &  \gamma_3
\end{array}
\end{bmatrix}
\Bigr(\frac{B_z}{2}\Bigl)\,.\label{eq:HBz}
\end{eqnarray}  
In high field, in practice above a few $\mu\mathrm{T}$, the Zeeman term dominates and the basis states approach the eigenstates of $H$.  In this limit the directly resolved transitions are governed by the diagonal scalar couplings  $J_{12}$ and $J_\Sigma$.
The LAC instead occurs when the diagonal energies of $\ket{\phi_a}$ and $\ket{\phi_b}$ become degenerate ($\braket{\phi_a| H |\phi_a}=\braket{\phi_b| H |\phi_b}$), where using \autoref{eq:bprimedef} we obtain
\begin{eqnarray}
H_{B_{\rm LAC}} &\sim&
-\begin{bmatrix}
\begin{array}{cc|c}
1 & 0 & 0 \\
0 & 1 & 0 \\\hline 
0 & 0 & 1
\end{array}
\end{bmatrix}\frac{\gamma_{\rm H} B_{\rm LAC}}{2} \label{eq:HLAC} \\ &&+
\begin{bmatrix}
\begin{array}{cc|c}
- 2J_{12} - J_{\Sigma} &  -2 \sqrt{2}J_\Delta & \cdot \\ 
-2 \sqrt{2}J_\Delta & - 2J_{12} - J_{\Sigma} & \cdot \\\hline
\cdot & \cdot & 6J_{12} - J_\Sigma
\end{array}
\end{bmatrix}
\Bigr(\frac{\pi}{4}\Bigl)
\nonumber 
\,. 
\end{eqnarray}
Elements marked ``$\cdot$'' are negligible at this field, because couplings to $\ket{\phi_c}$ are suppressed by the hierarchy $|J_{12}|\gg|J_\Sigma|,|J_\Delta|$. 
The LAC eigenstates are therefore well approximated by the symmetric and antisymmetric combinations of $\ket{\phi_a}$ and $\ket{\phi_b}$, while $\ket{\phi_c}$ remains comparatively isolated.

\textit{Appendix B. Liouville space description.---} 
Near the positive-field LAC, the pair of states $\{\ket{\phi_a},\ket{\phi_b}\}$ form a near-isolated two-level subspace, thus to a good approximation the corresponding density-operator dynamics are confined to a four-dimensional Liouville subspace,
$\mathcal{S}^{(ab)}=\{E,A_x,A_y,A_z\}$, 
comprising basis operators
\begin{eqnarray}
E &=& \ket{\phi_a}\bra{\phi_a}
+\ket{\phi_b}\bra{\phi_b}, \nonumber\\
A_x &=& \frac{1}{2}
\left(
\ket{\phi_a}\bra{\phi_b}
+\ket{\phi_b}\bra{\phi_a}
\right), \nonumber\\
A_y &=& \frac{i}{2}
\left(
\ket{\phi_a}\bra{\phi_b}
-\ket{\phi_b}\bra{\phi_a}
\right), \nonumber\\
A_z &=& \frac{1}{2}
\left(
\ket{\phi_a}\bra{\phi_a}
-\ket{\phi_b}\bra{\phi_b}
\right).
\label{eq:stopbasisdefn}
\end{eqnarray}
Here $E$ is the identity within the isolated subspace, $A_z$ is the population
imbalance between $\ket{\phi_a}$ and $\ket{\phi_b}$, and $A_x,A_y$ are the
corresponding coherences.  
These operators obey $[E, A_\xi]=0$ and $[A_j, A_k] = \mathrm{i}\,\epsilon_{jkl} A_l$ so the Liouville-space projection of the density operator, $\rho(t)$,
\begin{equation}
    \mathbf{\rho}^{(ab)}(t)
    =
    \left[
    (A_x|\rho(t)),
    (A_y|\rho(t)),
    (A_z|\rho(t))
    \right],
\end{equation}
where $(Q|\rho)\equiv {\rm Tr}(Q^\dagger\rho)$, evolves exactly as a Bloch vector under the
projected Hamiltonian, as given in \autoref{eq:3DsubspaceLvN}. 
Projecting $H$ onto the same operator basis gives the effective angular velocity, \autoref{eq:wlacvectormodel}.  
We note the norm of the projected vector (\autoref{eq:wlacvectormodel}) is the precession frequency $\omega_{\rm LAC}$, and for small $\delta B$ this reproduces the perturbative expression of \autoref{eq:wlac2ndorder}:
\begin{equation}
    |\mathbf{\omega}_{\rm LAC}|
    =
    \pi\sqrt{2}|J_\Delta|
    +
    \frac{
    (\gamma_{\rm H}-\gamma_{\rm C})^2\delta B^2
    }
    {
    2\pi\sqrt{2}|J_\Delta|
    }
    +
    O(\delta B^4)\,.
\end{equation}

The nuclear magnetization along a given laboratory-frame axis $\xi$ is given by $\braket{M_\xi}\equiv(M_\xi|\rho)$, where
\begin{equation}
    M_\xi=\sum_i\gamma_i I_{i\xi}\,.
\end{equation}
Transverse magnetization is always zero because of the zero-quantum nature of the subspace: $(Q|M_x) = (Q|M_y) =0$.  However, the $z$-axis component is nonzero, dependent on the population difference term: $A_z$: $(A_x|M_z)=(A_y|M_z)=0$ and $(A_z|M_z)=(\gamma_{\rm H}-\gamma_{\rm C})/2$. 
Thus, changes in the latitude of the Bloch vector generate changes in the observable magnetization along $B_z$.  Coherences $A_x$ and $A_y$ are not directly detected.

For ideal PHIP preparation, the initial density operator is \textsuperscript{1}H singlet order,
\begin{equation}
    \rho(0)
    =
    \ket{S_0\alpha}\bra{S_0\alpha}
    +
    \ket{S_0\beta}\bra{S_0\beta}
    \equiv
    \frac{1}{4}
    -
    \mathbf{I}_1\cdot\mathbf{I}_2 .
\end{equation}
The first term projects into the positive-field LAC manifold as
\begin{equation}
    \mathbf{\rho}^{(ab)}(0)=(0,0,1/2),
\end{equation}
while the second term gives the analogous contribution in the \(m_I=-1/2\) manifold. At the positive-field LAC, free evolution gives
\begin{equation}
    \rho^{(ab)}(\theta)
    =
    A_z\cos\theta
    +
    A_y\sin\theta
    +
    \frac{E}{2},
    \qquad
    \theta=\omega_{\rm LAC}t .
\end{equation}
Including the corresponding contribution from the \(m_I=-1/2\) manifold, the longitudinal magnetization becomes
\begin{eqnarray}
    \langle M_z\rangle
    &=&
    \frac{1}{2}
    \left(
    A_z\cos\theta
    +
    \frac{E}{2}
    +
    \ket{S_0\beta}\bra{S_0\beta}
    \middle| M_z
    \right) \nonumber\\
    &=&
    \frac{\gamma_{\rm H}-\gamma_{\rm C}}{4}
    \left(\cos\theta-1\right)\,,
\end{eqnarray}
or equivalently \autoref{eq:MzvsTime}.


\textit{Appendix C. Incoherent dynamics of \fumarateC at the LAC point.---}
Spin-singlet populations and singlet-triplet coherences often outlive conventional relaxation modes because they are partially immune to pair-correlated pathways.  In high magnetic fields, for example in \fumarateC at $\sim$\SI{10}{T}, population difference between the \textsuperscript{1}H singlet and triplet manifolds decays with an exponential time constant $T_S = 46 \pm 7$ s, substantially longer than $T_1 \sim \SI{23}{s}$ reported for the same system.\cite{ripka_hyperpolarized_2018}
Other studies report long $T_S$ values in comparable regimes. \cite{eills2021singlet}.

It is natural to ask whether such protection persists at the LAC field.  
A basic analysis suggests that \fumarateC at the LAC field does \textit{not} support long-lived coherences. The eigenstates $(\ket{\phi_a} \pm \ket{\phi_b})/\sqrt{2}$ each contain equal mixtures of singlet and triplet components, which eliminates the exchange symmetry that protects against correlated relaxation \cite{carravetta2005theory}.

A simple random-field model makes this explicit. Let the \textsuperscript{1}H spins experience zero-mean stochastic fields $\mathbf{B}_1$ and $\mathbf{B}_2$ with root-mean-square amplitude $B_{\rm rms}$ and cross-correlation $c_{12}=\braket{\mathbf{B}_1 \cdot \mathbf{B}_2}/B_{\rm rms}^2$. In the motional-narrowing limit, Redfield theory gives the auto-relaxation rates of the reduced subspace operators
\begin{eqnarray}
    \frac{1}{T_{A_x}} = \frac{1}{T_{A_y}} &=& \gamma_{\rm H}^2 B_{\rm rms}^2 \tau_c \,(5 - c_{12})/2\,, \\[3pt]
    \frac{1}{T_{A_z}} &=& 2\gamma_{\rm H}^2 B_{\rm rms}^2 \tau_c\,.
\end{eqnarray}
These rates become identical when $c_{12}=1$. For comparison, an isolated \textsuperscript{13}C nucleus should relax with
\begin{equation}
    \frac{1}{T_2(\text{\textsuperscript{13}C})} = 2\gamma_{\rm C}^2 B_{\rm rms}^2 \tau_c\,,
\end{equation}
which is slower by over an order of magnitude because $\gamma_{\rm C}^2 \approx 0.06 \gamma_{\rm H}^2$. Thus, operation at the LAC does not provide protection against homogeneous relaxation, and any advantage instead stems from lower sensitivity to inhomogeneity in $\delta B$.


\section*{Methods}




\textit{PHIP polarization of \fumarateC ---}
To produce hyperpolarized \fumarateC, the precursor solution for all experiments was 125\,mM acetylenedicarboxylic acid disodium salt (Merck), 125\,mM sodium sulfite, and 3.5\,mM ruthenium catalyst (pentamethyl cyclopentadienyl tris(acetonitrile) ruthenium(II) hexafluorophosphate) (Merck, CAS 99604-67-8) in $\sim$99$\%$ D\textsubscript{2}O, at pH 10-11.  For each measurement, we pipetted \SI{600}{\micro L} of the solution into a 5\,mm screw-top NMR tube, and the tube was sealed. The screw top had two 1/16'' (o.d.) PTFE capillaries passing through, acting as an inlet and an outlet for the hydrogen gas. We pressurized the tube to 5\,bar with parahydrogen, and placed the lower part of the tube into a mineral oil bath for the solution to warm (\SI{115}{\celsius}, 10\,s).  While still in the hot bath, we flowed parahydrogen through the capillaries to bubble through the solution (5\,bar for 60\,s). A color change of the solution from yellow to pink/red color indicated completion of the reaction.  We then depressurized the tube, cooled the sample in a water bath, opened the tube, and extracted the solution through a PTFE capillary into a 1\,mL syringe. The solution was then injected through a capillary tube into a 2\,mL, 8-425 vial (Merck) located inside the shielded apparatus of \autoref{fig:fig3}; the delay between the end of the reaction and the start of field sweeps was 15 to \SI{20}{s}.

\textit{Field control ---}
For magnetic field control and shielding we used a cylindrical multilayer MuMetal magnetic shield (Twinleaf model MS-1F, dimensions 30\,cm length and 25\,cm diameter), with a solenoid coil (7.5\,mT/A, 15\,mm diameter) positioned inside, parallel to the cylinder axis.  $B_z$ magnetic fields were generated using a 12-bit digital-to-analog converter (DAC) integrated circuit (MCP4822, MicroChip Technology Inc.) whose output was connected to the solenoid and a series shunt resistor.  The DAC was digitally interfaced with a microcontroller, provided as part of the ``NMRduino'' open-source magnetic resonance platform\cite{tayler2024nmrduino}.  

We calibrated the solenoid field strength at each DAC output voltage from the measured \textsuperscript{1}H precession frequencies of distilled water via a pulse-acquire experiment, following ex-situ prepolarization of the sample at \SI{2}{T}\cite{tayler2024nmrduino}.

\textit{Optical magnetometry ---}
The OPM used for detection comprised a cube-shaped alkali-metal vapor cell as its sensitive element ($5 \times 5 \times 8$ mm$^3$), which contained \textsuperscript{87}Rb and N\textsubscript{2} buffer gas heated to \SI{150}{\celsius}.  A D\textsubscript{1}-resonant (\SI{795}{nm}), circularly polarized light beam passed through one of the cube faces to optically pump the atomic spin polarization of the vapor.  A second, off-resonance beam passed through one of the other faces, and out of the opposing side, to probe the atomic polarization via optical rotation.  The noise floor in the 10 to \SI{100}{Hz} frequency range was \SI{12}{fT\per\sqrt{Hz}}. 

The remaining cube face was against the exterior of the solenoid coil, near to the sample vial, with a standoff distance of 5-6 mm between the vapor cell and the vial.  Further details of the magnetometer can be found in past work\cite{Bodenstedt2021natcomm}.

\textit{Data acquisition and processing ---}
We used a balanced polarimeter (Thorlabs PDB210A) to detect the rotation angle of the OPM probe beam, which produced a differential output voltage linearly proportional to the $z$- and $x$-axis magnetic field amplitude.  The voltage was digitized by a second ``NMRduino''\cite{tayler2024nmrduino} device (16 bit, $\sim$\SI{150}{\micro\volt\per bit}, 1--5 kHz sampling rate), and streamed to a computer via USB and displayed on-screen in real time after digital low-pass filtering.  The raw signal data were also stored in a file on the computer for future retrieval, processing and plotting.  Data processing operations to generate the plots shown in \autoref{fig:fig3} involved simple operations in \textsc{Mathematica} (Wolfram Inc.) including: (i) a \SI{25}{Hz} second-order low-pass Hamming or moving-average filter to suppress 50 Hz mains-electricity noise, which was the dominant noise source of the magnetometer, and (ii) Fourier transformation.  A record of these operations is contained in the Supporting Dataset.

\textit{Simulations ---}
We performed simulations using the \textsc{SpinDynamica} packages for \textsc{Mathematica}\cite{bengs2018spindynamica}.  All simulations involved three spins, two \textsuperscript{1}H and one \textsuperscript{13}C, and the interactions considered were the Zeeman interaction between the spins, the applied magnetic fields (in $x$, $y$, or $z$) and the spin-spin $J$\nobreakdash-couplings (untruncated couplings in all cases). Chemical shifts were omitted because at such low fields these terms are negligible.

\section*{Acknowledgments}
\paragraph*{Funding:}
The work described is funded by: 
the Spanish Ministry of Science MCIN with funding from European Union NextGenerationEU (PRTR-C17.I1) and by Generalitat de Catalunya ``Severo Ochoa'' Center of Excellence CEX2019-000910-S; 
the Spanish Ministry of Science projects 
DHYMOND (PID2024-160223OB-I00), 
AMAREI (CNS2023-144106),
SEE-13-MRI (CPP2022-009771), 
SAPONARIA (PID2021-123813NB-I00) plus 
RYC2020-029099-I and RYC2022-035450-I, funded by MCIN/AEI /10.13039/501100011033; 
Generalitat de Catalunya through the CERCA program;  
Ag\`{e}ncia de Gesti\'{o} d'Ajuts Universitaris i de Recerca Grant No. 2021 FI\_B\_01039; 
Fundaci\'{o} Privada Cellex; 
Fundaci\'{o} Mir-Puig;
and the BIST--``la Caixa'' initiative in Chemical Biology (CHEMBIO).
The project has received funding from the European Union's Horizon 2020 Research and Innovation Programme under the Marie Sk\l{}odowska-Curie Grant Agreement 101063517 and also the European Commission project Field-SEER (ERC 101097313).
This work was supported by the Initiative and Networking Fund of the Helmholtz Association (Project No. VH-NG-20-20).



\paragraph*{Competing interests:} There are no competing interests to declare.

\paragraph*{Data availability statement:}
All data and code needed to evaluate and reproduce the results in the paper are publicly accessible via the web at \href{https://doi.org/10.5281/zenodo.19361842}{https://doi.org/10.5281/zenodo.19361842}.

\bibliography{references} 

@article{Bodenstedt2021natcomm,
  url = {https://dx.doi.org/10.1038/s41467-021-24248-9},
  year = {2021},
  pages = {4041},
  month = jun,
  publisher = {Springer Science and Business Media {LLC}},
  volume = {12},
  number = {1},
  author = {Sven Bodenstedt and Morgan W. Mitchell and Michael C. D. Tayler},
  title = {{Fast-field-cycling Ultralow-field Nuclear Magnetic Relaxation Dispersion}},
  journal = {Nat. Commun.}
}

@article{tayler2024nmrduino,
  title = {{NMRduino: A Modular, Open-Source, Low-Field Magnetic Resonance Platform}},
  volume = {362},
  ISSN = {1090-7807},
  url = {http://dx.doi.org/10.1016/j.jmr.2024.107665},
  journal = {J. Magn. Reson.},
  publisher = {Elsevier BV},
  author = {Tayler,  Michael C. D. and Bodenstedt,  Sven},
  year = {2024},
  month = may,
  pages = {107665}
}

@article{johannesson2004transfer,
  title={{Transfer of Para-Hydrogen Spin Order into Polarization by Diabatic Field Cycling}},
  author={J{\'o}hannesson, Haukur and Axelsson, Oskar and Karlsson, Magnus},
  journal={C. R. Phys.},
  volume={5},
  number={3},
  pages={315--324},
  year={2004},
  url = {https://dx.doi.org/10.1016/j.crhy.2004.02.001},
  publisher={Elsevier}
}

@article{rodin2021constant,
  title={{Constant-Adiabaticity Ultralow Magnetic Field Manipulations of Parahydrogen-Induced Polarization: Application to an {AA'X} Spin System}},
  author={Rodin, Bogdan A and Eills, James and Picazo-Frutos, Rom{\'a}n and Sheberstov, Kirill F and Budker, Dmitry and Ivanov, Konstantin L},
  journal={Phys. Chem. Chem. Phys.},
  volume={23},
  number={12},
  pages={7125--7134},
  year={2021},
  url = {https://dx.doi.org/10.1039/d0cp06581a},
  publisher={Royal Society of Chemistry}
}

@article{eills2023spin,
author = {Eills, James and Budker, Dmitry and Cavagnero, Silvia and Chekmenev, Eduard Y. and Elliott, Stuart J. and Jannin, Sami and Lesage, Anne and Matysik, J\"{o}rg and Meersmann, Thomas and Prisner, Thomas and Reimer, Jeffrey A. and Yang, Hanming and Koptyug, Igor V.},
title = {{Spin Hyperpolarization in Modern Magnetic Resonance}},
journal = {Chem. Rev.},
volume = {123},
issue = {4},
pages = {1417--1551},
url={https://dx.doi.org/10.1021/acs.chemrev.2c00534},
year = {2023}}

@article{mouloudakis2023real,
  title={{Real-Time Polarimetry of Hyperpolarized \textsuperscript{13}{C} Nuclear Spins Using an Atomic Magnetometer}},
  author={Mouloudakis, Kostas and Bodenstedt, Sven and Azagra, Marc and Mitchell, Morgan W and Marco-Rius, Irene and Tayler, Michael C. D.},
  journal={J. Chem. Phys. Lett.},
  volume={14},
  number={5},
  pages={1192--1197},
  year={2023},
  url = {https://dx.doi.org/10.1021/acs.jpclett.2c03864},
  publisher={ACS Publications}
}

@article{levitt1996demagnetization,
  title = {{Demagnetization Field Effects in Two-Dimensional Solution NMR}},
  author = {Levitt, Malcolm H},
  journal={Concepts Magn. Reson.},
  volume={8},
  number={2},
  pages={77--103},
  year={1996},
  url = {https://dx.doi.org/10.1039/d1cp04653e},
  publisher={Wiley Online Library}
}

@article{bengs2018spindynamica,
  title={{SpinDynamica: Symbolic and Numerical Magnetic Resonance in a Mathematica Environment}},
  author={Bengs, Christian and Levitt, Malcolm H},
  journal={Magn. Reson. Chem.},
  volume={56},
  number={6},
  pages={374--414},
  year={2018},
  url = {https://dx.doi.org/10.1002/mrc.4642},
  publisher={Wiley Online Library}
}

@article{dagys2024robust,
  title={{Robust Parahydrogen-Induced Polarization at High Concentrations}},
  volume = {10},
  ISSN = {2375-2548},
  url = {http://dx.doi.org/10.1126/sciadv.ado0373},
  number = {30},
  journal = {Sci. Adv.},
  publisher = {American Association for the Advancement of Science (AAAS)},
  author = {Dagys,  Laurynas and Korzeczek,  Martin C. and Parker,  Anna J. and Eills,  James and Blanchard,  John W. and Bengs,  Christian and Levitt,  Malcolm H. and Knecht,  Stephan and Schwartz,  Ilai and Plenio,  Martin B.},
  year = {2024},
  month = jul 
}

@article{bax_enhanced_1980,
	title = {{Enhanced NMR Resolution by Restricting the Effective Sample Volume}},
	volume = {37},
	copyright = {https://www.elsevier.com/tdm/userlicense/1.0/},
	issn = {00222364},
	url = {https://dx.doi.org/10.1016/0022-2364(80)90104-3},
	number = {1},
	urldate = {2025-06-28},
	journal = {J. Magn. Reson. (1969)},
	author = {Bax, Ad and Freeman, Ray},
	month = jan,
	year = {1980},
	pages = {177--181},
}

@article{eills_live_2024,
  title = {{Live Magnetic Observation of Parahydrogen Hyperpolarization Dynamics}},
  volume = {121},
  issn = {0027-8424, 1091-6490},
  url = {https://dx.doi.org/10.1073/pnas.2410209121},
  number = {43},
  journal = {Proc. Natl. Acad. Sci. U.S.A.},
  author = {Eills, James and Mitchell, Morgan W. and Rius, Irene Marco and Tayler, Michael C. D.},
  month = oct,
  year = {2024},
  pages = {e2410209121},
}

@article{eck_level_1967,
	title = {Level crossings and anticrossings},
	volume = {33},
	copyright = {https://www.elsevier.com/tdm/userlicense/1.0/},
	issn = {00318914},
	url = {https://dx.doi.org/10.1016/0031-8914(67)90266-2},
	number = {1},
	urldate = {2025-06-28},
	journal = {Physica},
	author = {Eck, T.G.},
	month = jan,
	year = {1967},
	pages = {157--162},
}

@article{heil_spin_2013,
	title = {{Spin Clocks: Probing Fundamental Symmetries in Nature}},
	volume = {525},
	copyright = {© 2013 by WILEY-VCH Verlag GmbH \& Co. KGaA, Weinheim},
	issn = {1521-3889},
	shorttitle = {Spin clocks},
	url = {https://dx.doi.org/10.1002/andp.201300048},
	number = {8-9},
	urldate = {2025-07-31},
	journal = {Ann. Phys.},
	author = {Heil, Werner and Gemmel, Claudia and Karpuk, Sergei and Sobolev, Yuri and Tullney, Kathlynne and Allmendinger, Fabian and Schmidt, Ulrich and Burghoff, Martin and Kilian, Wolfgang and Knappe-Gr{\"u}neberg, Silvia and Schnabel, Allard and Seifert, Frank and Trahms, Lutz},
	year = {2013},
	pages = {539--549},
}

@article{Ledbetter2012PRLcomagnetometer,
  title = {Liquid-State Nuclear Spin Comagnetometers},
  volume = {108},
  ISSN = {1079-7114},
  url = {http://doi.org/10.1103/PhysRevLett.108.243001},
  DOI = {10.1103/physrevlett.108.243001},
  number = {24},
  journal = {Physical Review Letters},
  publisher = {American Physical Society (APS)},
  author = {Ledbetter,  M. P. and Pustelny,  S. and Budker,  D. and Romalis,  M. V. and Blanchard,  J. W. and Pines,  A.},
  year = {2012},
  pages = {243001},
  month = jun
}

@article{Wu2018comagnetometer,
  title = {Nuclear-Spin Comagnetometer Based on a Liquid of Identical Molecules},
  volume = {121},
  ISSN = {1079-7114},
  url = {http://doi.org/10.1103/PhysRevLett.121.023202},
  DOI = {10.1103/physrevlett.121.023202},
  number = {2},
  journal = {Physical Review Letters},
  publisher = {American Physical Society (APS)},
  author = {Wu,  Teng and Blanchard,  John W. and Jackson Kimball,  Derek F. and Jiang,  Min and Budker,  Dmitry},
  year = {2018},
  pages = {023202},
  month = jul 
}

@article{ripka_hyperpolarized_2018,
	title = {Hyperpolarized Fumarate via Parahydrogen},
	volume = {54},
	issn = {1364-548X},
	url = {https://dx.doi.org/10.1039/C8CC06636A},
	number = {86},
	urldate = {2020-04-09},
	journal = {Chem. Commun.},
	author = {Ripka, Barbara and Eills, James and Kou{\v r}ilov{\'a}, Hana and Leutzsch, Markus and Levitt, Malcolm H. and M{\"u}nnemann, Kerstin},
	month = oct,
	year = {2018},
	pages = {12246--12249},
}

@article{eills_real-time_2019,
	title = {Real-{Time} {Nuclear} {Magnetic} {Resonance} {Detection} of {Fumarase} {Activity} {Using} {Parahydrogen}-{Hyperpolarized} {[1-$^{13}$C]Fumarate}},
	volume = {141},
	issn = {0002-7863},
	url = {https://dx.doi.org/10.1021/jacs.9b10094},
	number = {51},
	urldate = {2025-07-30},
	journal = {J. Am. Chem. Soc.},
	author = {Eills, James and Cavallari, Eleonora and Carrera, Carla and Budker, Dmitry and Aime, Silvio and Reineri, Francesca},
	month = dec,
	year = {2019},
	pages = {20209--20214},
}

@article{gallagher_production_2009,
	title = {{Production of Hyperpolarized {[1,4-$^{13}$C$_2$]}malate from {[1,4-$^{13}$C$_2$]}fumarate is a Marker of Cell Necrosis and Treatment Response in Tumors}},
	volume = {106},
	url = {https://dx.doi.org/10.1073/pnas.0911447106},
	number = {47},
	urldate = {2025-07-30},
	journal = {Proc. Natl. Acad. Sci. U.S.A.},
	author = {Gallagher, Ferdia A. and Kettunen, Mikko I. and Hu, De-En and Jensen, Pernille R. and Zandt, {René in ‘t} and Karlsson, Magnus and Gisselsson, Anna and Nelson, Sarah K. and Witney, Timothy H. and Bohndiek, Sarah E. and Hansson, Georg and Peitersen, Torben and Lerche, Mathilde H. and Brindle, Kevin M.},
	month = nov,
	year = {2009},
	pages = {19801--19806},
}

@article{gierse_parahydrogen-polarized_2023,
	title = {Parahydrogen-{Polarized} {Fumarate} for {Preclinical} in {Vivo} {Metabolic} {Magnetic} {Resonance} {Imaging}},
	volume = {145},
	issn = {0002-7863},
	url = {https://dx.doi.org/10.1021/jacs.2c13830},
	number = {10},
	urldate = {2025-07-30},
	journal = {J. Am. Chem. Soc.},
	author = {Gierse, Martin and Nagel, Luca and Keim, Michael and Lucas, Sebastian and Speidel, Tobias and Lobmeyer, Tobias and Winter, Gordon and Josten, Felix and Karaali, Senay and Fellermann, Maximilian and Scheuer, Jochen and M{\"u}ller, Christoph and van Heijster, Frits and Skinner, Jason and L{\"o}ffler, Jessica and Parker, Anna and Handwerker, Jonas and Marshall, Alastair and Salhov, Alon and El-Kassem, Bilal and Vassiliou, Christophoros and Blanchard, John W. and Picazo-Frutos, Román and Eills, James and Barth, Holger and Jelezko, Fedor and Rasche, Volker and Schilling, Franz and Schwartz, Ilai and Knecht, Stephan},
	month = mar,
	year = {2023},
	pages = {5960--5969},
}

@article{vathyam_homogeneous_1996,
	title = {Homogeneous {NMR} {Spectra} in {Inhomogeneous} {Fields}},
	volume = {272},
	url = {https://dx.doi.org/10.1126/science.272.5258.92},
	number = {5258},
	urldate = {2025-07-30},
	journal = {Science},
	author = {Vathyam, Sujatha and Lee, Sanghyuk and Warren, Warren S.},
	month = apr,
	year = {1996},
	pages = {92--96},
}

@article{munowitz_multiple-quantum_1986,
	title = {Multiple-{Quantum} {Nuclear} {Magnetic} {Resonance} {Spectroscopy}},
	volume = {233},
	url = {https://dx.doi.org/10.1126/science.233.4763.525},
	number = {4763},
	urldate = {2025-07-30},
	journal = {Science},
	author = {Munowitz, M. and Pines, A.},
	month = aug,
	year = {1986},
	pages = {525--531},
}

@article{aue_twodimensional_1976,
	title = {Two-Dimensional Spectroscopy. Application to Nuclear Magnetic Resonance},
	volume = {64},
	issn = {0021-9606},
	url = {https://dx.doi.org/10.1063/1.432450},
	number = {5},
	urldate = {2025-07-30},
	journal = {J. Chem. Phys.},
	author = {Aue, W. P. and Bartholdi, E. and Ernst, R. R.},
	month = mar,
	year = {1976},
	pages = {2229--2246},
}

@article{lacey_high-resolution_1999,
	title = {High-{Resolution} {NMR} {Spectroscopy} of {Sample} {Volumes} from 1 {nL} to 10 $\mu${L}},
	volume = {99},
	issn = {0009-2665, 1520-6890},
	url = {https://dx.doi.org/10.1021/cr980140f},
	number = {10},
	urldate = {2025-08-26},
	journal = {Chem. Rev.},
	author = {Lacey, Michael E. and Subramanian, Raju and Olson, Dean L. and Webb, Andrew G. and Sweedler, Jonathan V.},
	month = oct,
	year = {1999},
	pages = {3133--3152},
}

@article{cai_high-resolution_2009,
	title = {High-{Resolution} {Solution} {NMR} {Spectra} in {Inhomogeneous} {Magnetic} {Fields}},
	volume = {5},
	issn = {15734110},
	url = {https://dx.doi.org/10.2174/157341109787047844},
	number = {1},
	urldate = {2025-08-26},
	journal = {Curr. Anal. Chem.},
	author = {Cai, Shuhui and Zhang, Wen and Chen, Zhong},
	month = jan,
	year = {2009},
	pages = {70--83},
}

@article{pelupessy_high-resolution_2009,
	title = {High-{Resolution} {NMR} in {Magnetic} {Fields} with {Unknown} {Spatiotemporal} {Variations}},
	volume = {324},
	issn = {0036-8075, 1095-9203},
	url = {https://dx.doi.org/10.1126/science.1175102},
	number = {5935},
	urldate = {2025-08-26},
	journal = {Science},
	author = {Pelupessy, Philippe and Rennella, Enrico and Bodenhausen, Geoffrey},
	month = jun,
	year = {2009},
	pages = {1693--1697},
}

@article{terenzi_enabling_2019,
	title = {Enabling {High} {Spectral} {Resolution} of {Liquid} {Mixtures} in {Porous} {Media} by {Antidiagonal} {Projections} of {Two}-{Dimensional}$^{\textrm{1}}$ {H} {NMR} {COSY} {Spectra}},
	volume = {10},
	copyright = {https://doi.org/10.15223/policy-029},
	issn = {1948-7185, 1948-7185},
	url = {https://dx.doi.org/10.1021/acs.jpclett.9b02334},
	number = {19},
	urldate = {2025-08-26},
	journal = {J. Phys. Chem. Lett.},
	author = {Terenzi, Camilla and Sederman, Andrew J. and Mantle, Michael D. and Gladden, Lynn F.},
	month = oct,
	year = {2019},
	pages = {5781--5785},
}

@article{eills2021singlet,
  title = {Singlet-Contrast Magnetic Resonance Imaging: Unlocking Hyperpolarization with Metabolism},
  volume = {60},
  ISSN = {1521-3773},
  url = {http://dx.doi.org/10.1002/anie.202014933},
  number = {12},
  journal = {Angew. Chem. Intl. Ed.},
  publisher = {Wiley},
  author = {Eills,  J. and Cavallari,  E. and Kircher,  R. and Di Matteo,  G. and Carrera,  C. and Dagys,  L. and Levitt,  M. H. and Ivanov,  K. L. and Aime,  S. and Reineri,  F. and M{\''u}nnemann,  K. and Budker,  D. and Buntkowsky,  G. and Knecht,  S.},
  year = {2021},
  month = feb,
  pages = {6791--6798}
}

@article{carravetta2005theory,
  title = {{Theory of Long-Lived Nuclear Spin States in Solution Nuclear Magnetic Resonance. I. Singlet States in Low Magnetic Field}},
  volume = {122},
  ISSN = {1089-7690},
  url = {http://dx.doi.org/10.1063/1.1893983},
  number = {21},
  journal = {J. Chem. Phys.},
  publisher = {AIP Publishing},
  author = {Carravetta,  Marina and Levitt,  Malcolm H.},
  year = {2005},
  month = jun 
}

@article{wang2025nuclear,
  title = {{Nuclear Spins in a Solid Exceeding 10-Hour Coherence Times for Ultra-Long-Term Quantum Storage}},
  volume = {6},
  ISSN = {2691-3399},
  url = {http://dx.doi.org/10.1103/PRXQuantum.6.010302},
  number = {1},
  journal = {PRX Quantum},
  publisher = {American Physical Society (APS)},
  author = {Wang,  Fudong and Ren,  Miaomiao and Sun,  Weiye and Guo,  Mucheng and Sellars,  Matthew J. and Ahlefeldt,  Rose L. and Bartholomew,  John G. and Yao,  Juan and Liu,  Shuping and Zhong,  Manjin},
  year = {2025},
  month = jan 
}

@article{manucharyan2009fluxonium,
  title = {{Fluxonium: Single Cooper-Pair Circuit Free of Charge Offsets}},
  volume = {326},
  ISSN = {1095-9203},
  url = {http://dx.doi.org/10.1126/science.1175552},
  number = {5949},
  journal = {Science},
  publisher = {American Association for the Advancement of Science (AAAS)},
  author = {Manucharyan,  Vladimir E. and Koch,  Jens and Glazman,  Leonid I. and Devoret,  Michel H.},
  year = {2009},
  month = oct,
  pages = {113--116}
}

@article{bao2022fluxonium,
  title = {{Fluxonium: An Alternative Qubit Platform for High-Fidelity Operations}},
  volume = {129},
  ISSN = {1079-7114},
  url = {http://dx.doi.org/10.1103/PhysRevLett.129.010502},
  number = {1},
  journal = {Phys. Rev. Lett.},
  publisher = {American Physical Society (APS)},
  author = {Bao,  Feng and Deng,  Hao and Ding,  Dawei and Gao,  Ran and Gao,  Xun and Huang,  Cupjin and Jiang,  Xun and Ku,  Hsiang-Sheng and Li,  Zhisheng and Ma,  Xizheng and Ni,  Xiaotong and Qin,  Jin and Song,  Zhijun and Sun,  Hantao and Tang,  Chengchun and Wang,  Tenghui and Wu,  Feng and Xia,  Tian and Yu,  Wenlong and Zhang,  Fang and Zhang,  Gengyan and Zhang,  Xiaohang and Zhou,  Jingwei and Zhu,  Xing and Shi,  Yaoyun and Chen,  Jianxin and Zhao,  Hui-Hai and Deng,  Chunqing},
  year = {2022},
  month = jun 
}

@article{ivanov2014role,
  title = {{The Role of Level Anti-Crossings in Nuclear Spin Hyperpolarization}},
  volume = {81},
  ISSN = {0079-6565},
  url = {http://dx.doi.org/10.1016/j.pnmrs.2014.06.001},
  journal = {Prog. Nucl. Magn. Reson. Spectrosc.},
  publisher = {Elsevier BV},
  author = {Ivanov,  Konstantin L. and Pravdivtsev,  Andrey N. and Yurkovskaya,  Alexandra V. and Vieth,  Hans-Martin and Kaptein,  Robert},
  year = {2014},
  month = aug,
  pages = {1--36}
}

@article{buljubasich2012level,
  title = {{Level Anti-Crossings in ParaHydrogen Induced Polarization Experiments with Cs-Symmetric Molecules}},
  volume = {219},
  ISSN = {1090-7807},
  url = {http://dx.doi.org/10.1016/j.jmr.2012.03.020},
  journal = {J. Magn. Reson.},
  publisher = {Elsevier BV},
  author = {Buljubasich,  L. and Franzoni,  M.B. and Spiess,  H.W. and M{\''u}nnemann,  K.},
  year = {2012},
  month = jun,
  pages = {33--40}
}

@article{broadway2016anticrossing,
  title = {{Anticrossing Spin Dynamics of Diamond Nitrogen-Vacancy Centers and All-Optical Low-Frequency Magnetometry}},
  volume = {6},
  ISSN = {2331-7019},
  url = {http://dx.doi.org/10.1103/PhysRevApplied.6.064001},
  number = {6},
  journal = {Phys. Rev. Appl.},
  publisher = {American Physical Society (APS)},
  author = {Broadway,  David A. and Wood,  James D. A. and Hall,  Liam T. and Stacey,  Alastair and Markham,  Matthew and Simpson,  David A. and Tetienne,  Jean-Philippe and Hollenberg,  Lloyd C. L.},
  year = {2016},
  month = dec 
}

@article{theis2015microtesla,
  title = {{Microtesla SABRE Enables 10\% Nitrogen-15 Nuclear Spin Polarization}},
  volume = {137},
  ISSN = {1520-5126},
  url = {http://dx.doi.org/10.1021/ja512242d},
  number = {4},
  journal = {J. Am. Chem. Soc.},
  publisher = {American Chemical Society (ACS)},
  author = {Theis,  Thomas and Truong,  Milton L. and Coffey,  Aaron M. and Shchepin,  Roman V. and Waddell,  Kevin W. and Shi,  Fan and Goodson,  Boyd M. and Warren,  Warren S. and Chekmenev,  Eduard Y.},
  year = {2015},
  month = jan,
  pages = {1404–1407}
}

@article{krishnan2013radiation,
  title = {{Radiation damping in modern NMR experiments: Progress and challenges}},
  volume = {68},
  ISSN = {0079-6565},
  url = {http://dx.doi.org/10.1016/j.pnmrs.2012.06.001},
  journal = {Prog. Nucl. Magn. Reson. Spectr.},
  publisher = {Elsevier BV},
  author = {Krishnan,  V.V. and Murali,  Nagarajan},
  year = {2013},
  month = jan,
  pages = {41–57}
}

@article{ledbetter2012liquid,
  title = {Liquid-State Nuclear Spin Comagnetometers},
  volume = {108},
  ISSN = {1079-7114},
  url = {http://doi.org/10.1103/PhysRevLett.108.243001},
  number = {24},
  journal = {Phys. Rev. Lett.},
  publisher = {American Physical Society (APS)},
  author = {Ledbetter,  M. P. and Pustelny,  S. and Budker,  D. and Romalis,  M. V. and Blanchard,  J. W. and Pines,  A.},
  year = {2012},
  pages={243001},
  month = jun 
}

@article{Warren1993,
  title = {{Generation of Impossible Cross-Peaks Between Bulk Water and Biomolecules in Solution NMR}},
  volume = {262},
  ISSN = {1095-9203},
  url = {http://dx.doi.org/10.1126/science.8266096},
  DOI = {10.1126/science.8266096},
  number = {5142},
  journal = {Science},
  publisher = {American Association for the Advancement of Science (AAAS)},
  author = {Warren,  Warren S. and Richter,  Wolfgang and Andreotti,  Amy Hamilton and Farmer,  Bennett T.},
  year = {1993},
  month = Dec,
  pages = {2005–2009}
}

@article{Baudin2007,
  title = {{Nonlinear NMR dynamics in hyperpolarized liquid 3He}},
  volume = {11},
  ISSN = {1878-1543},
  url = {http://dx.doi.org/10.1016/j.crci.2007.07.005},
  DOI = {10.1016/j.crci.2007.07.005},
  number = {4-5},
  journal = {Comptes Rendus. Chimie},
  publisher = {MathDoc/Centre Mersenne},
  author = {Baudin,  Emmanuel and Hayden,  Michael E. and Tastevin,  Geneviève and Nacher,  Pierre-Jean},
  year = {2007},
  month = Nov,
  pages = {560–567}
}

@article{Chu2020,
  author = {Chu, P.-H. and Kim, Y. J. and Savukov, I. M.},
  title = {Search for exotic spin-dependent interactions using polarized helium},
  journal = {arXiv},
  year = {2020},
  eprint = {2002.02495},
  archivePrefix = {arXiv},
  primaryClass = {physics.atom-ph}
}

@article{Arvanitaki2014,
  title = {{Resonantly Detecting Axion-Mediated Forces with Nuclear Magnetic Resonance}},
  volume = {113},
  ISSN = {1079-7114},
  url = {http://dx.doi.org/10.1103/PhysRevLett.113.161801},
  DOI = {10.1103/physrevlett.113.161801},
  number = {16},
  journal = {Phys. Rev. Lett.},
  publisher = {American Physical Society (APS)},
  author = {Arvanitaki,  Asimina and Geraci,  Andrew A.},
  year = {2014},
  pages = {161801},
  month = Oct 
}

@article{Ledbetter2002,
  title = {{Nonlinear Effects from Dipolar Interactions in Hyperpolarized Liquid ${129}$Xe}},
  volume = {89},
  ISSN = {1079-7114},
  url = {http://dx.doi.org/10.1103/PhysRevLett.89.287601},
  DOI = {10.1103/physrevlett.89.287601},
  number = {28},
  journal = {Phys. Rev. Lett.},
  publisher = {American Physical Society (APS)},
  author = {Ledbetter,  M. P. and Romalis,  M. V.},
  year = {2002},
  pages={287601},
  month = Dec 
}

@article{Sahin2022,
  title = {{High field magnetometry with hyperpolarized nuclear spins}},
  volume = {13},
  ISSN = {2041-1723},
  url = {http://dx.doi.org/10.1038/s41467-022-32907-8},
  DOI = {10.1038/s41467-022-32907-8},
  number = {1},
  journal = {Nat. Commun.},
  publisher = {Springer Science and Business Media LLC},
  author = {Sahin,  Ozgur and de Leon Sanchez,  Erica and Conti,  Sophie and Akkiraju,  Amala and Reshetikhin,  Paul and Druga,  Emanuel and Aggarwal,  Aakriti and Gilbert,  Benjamin and Bhave,  Sunil and Ajoy,  Ashok},
  year = {2022},
  month = sep 
}

@article{Garcon2019,
  title = {{Constraints on bosonic dark matter from ultralow-field nuclear magnetic resonance}},
  volume = {5},
  ISSN = {2375-2548},
  url = {http://dx.doi.org/10.1126/sciadv.aax4539},
  DOI = {10.1126/sciadv.aax4539},
  number = {10},
  journal = {Sci. Adv.},
  publisher = {American Association for the Advancement of Science (AAAS)},
  author = {Garcon,  Antoine and Blanchard,  John W. and Centers,  Gary P. and Figueroa,  Nataniel L. and Graham,  Peter W. and Jackson Kimball,  Derek F. and Rajendran,  Surjeet and Sushkov,  Alexander O. and Stadnik,  Yevgeny V. and Wickenbrock,  Arne and Wu,  Teng and Budker,  Dmitry},
  year = {2019},
  month = Oct 
}

@article{Limes2018,
  title = {{$^3$He-$^{129}$Xe Comagnetometery using $^{87}$Rb Detection and Decoupling}},
  volume = {120},
  ISSN = {1079-7114},
  url = {http://dx.doi.org/10.1103/PhysRevLett.120.033401},
  DOI = {10.1103/physrevlett.120.033401},
  number = {3},
  journal = {Phys. Rev. Lett.},
  publisher = {American Physical Society (APS)},
  author = {Limes,  M.E. and Sheng,  D. and Romalis,  M.V.},
  year = {2018},
  pages={033401},
  month = Jan 
}

\end{document}